\documentclass[twocolumn,showpacs,showkeys,preprintnumbers,amsmath,amssymb]{revtex4}

\usepackage{graphicx}
\usepackage{dcolumn}
\usepackage{bm}
\usepackage{color}

\begin{document}

\title{Gauge angle dependence in TDHFB calculations 
of ${}^{20}$O + ${}^{20}$O head-on collisions with the Gogny interaction 
}

\author{Yukio Hashimoto}                     
\email{hashimoto.yukio.gb@u.tsukuba.ac.jp}
\affiliation{Center for Computational Sciences, 
University of Tsukuba, Tsukuba 305-8571, Japan}

\author{Guillaume Scamps}
\email{scamps@nucl.phys.tohoku.ac.jp}
\affiliation{Department of Physics, Tohoku University, Sendai 980-8578, Japan}

\date{\today}

\begin{abstract}
A numerical method to solve the TDHFB equations by using a hybrid basis 
of the two-dimensional harmonic oscillator eigenfunctions 
and one-dimensional Lagrange mesh with the Gogny effective interaction 
is applied to the head-on collisions of the superfluid nuclei ${}^{20}$O's. 
Taking the energies around the barrier top 
energy, the trajectories, pairing energies, and numbers of transferred nucleons 
are displayed. Their dependence on the relative gauge angle at the initial time 
is studied by taking typical sample points of the gauge angle. 
It turned out that the functional form of the flux 
of the neutrons across a section plane 
is proportional to the sine of the two times of the gauge angle.
\end{abstract}

\maketitle

\section{Introduction}
\label{intro}
The nuclear superfluidity has attracted continuous attention of the 
nuclear physicists for more than fifty years. 
The structure of the nuclear ground states, the reaction mechanism 
of a pair of colliding nuclei, the dynamical properties of the 
fission processes, and so on, have been studied in connection with 
the pairing correlations among the nucleons in the nucleus (nuclei). 

The time-dependent mean-field theory has played the central role 
in the nuclear physics, and the time-dependent Hartree-Fock (TDHF) 
method is the foremost example of the time-dependent mean-field methods.
The TDHF has been widely in use in the investigations of the 
small-amplitude collective vibrations around the ground states 
as well as the large-amplitude collective motions 
in the nuclear fusion/fission processes~\cite{BKN,Negele,Maruhn,Umar,LinRes,Umar2,washiyama_PRC78,Umar3,Kedziora,Simenel14}.  
The TDHF has been extended to deal with the effects 
of the pairing correlation into the time-dependent Hartree-Fock Bogoliubov 
(TDHFB) or TDHF+BCS methods~\cite{YH-KN,Avez,Ebata,Stetcu,YH-EPJA,YH-PRC88,Scamps1}. 
 
In relation with the reaction processes of two superfluid nuclei, 
one of the long-standing, interesting subjects is to make clear 
the mechanism of the particle transfer phenomena 
influenced by the pairing correlation in the sub-barrier energy region.
The particle transfer process between the two superfluid nuclei 
might be expected to be analogous to the {\it Josephson effect} 
in the solid state physics~\cite{Joseph-1,Joseph-2}. 
In the Josephson effect, the electric current flows between 
the two superconducting objects separated by a thin insulator 
in proportion to the sine of the difference of the phases of the 
superconducting objects.

In the mean-field framework of the BCS approximation or Hartree-Fock 
Bogoliubov (HFB) method, the ground state of the superfluid nucleus 
is accompanied by a time-dependent phase (gauge angle) 
whose angular velocity is the chemical potential. 
The phase has no effects on the ground state property of the nucleus, 
or on its evolution with the TDHFB equations of motion.
In the collision process of two superfluid nuclei, on the other hand, 
the phase of the one nucleus could be different from that 
of the other's. 
We have no information about the effects of the difference of the phases 
(relative phase) of the two nuclei at the initial time on the 
dynamical property of the colliding nuclei later.

Recently, several groups have calculated the numbers 
of the transferred nucleons in the nuclear collisions 
by making use of the quantum mechanical method of the 
number projection, in which the angular variables are used as the 
generator coordinates (integration variables)~\cite{Simenel1,Scamps1,Sekizawa1,Sekizawa2}.  
In the framework of the mean-field calculations of the TDHFB, 
on the other hand, what is necessary to be made clear 
is the effect of the initial relative phase 
on the physical quantities such as trajectories, pairing energies, 
potential profiles, and so on, in the collision processes.
 
In this article, we report the first results of the application of the 
method of solving the TDHFB equations by using the hybrid basis of 
the two-dimensional harmonic oscillator eigenfunctions and one-dimensional 
Lagrange mesh~\cite{YH-PRC88} to the head-on collision processes of the superfluid oxygens 
${}^{20}$O's. 
Taking the boost energies so that the total energies are around the 
barrier top energy, we show the trajectories, variations of the pairing energies, 
and the number of the transferred nucleons with respect to the boost energies. 

In relation with the setting up of the initial conditions of the collision processes, 
we take four representative points (0, 45, 90, and 135 degrees) 
of the gauge angle with the purpose of 
studying the effects of them on the properties of the colliding nuclei.
The dependence of the potentials between the colliding nuclei, 
trajectories, and pairing energies on the relative phase are discussed.

This article is composed of the following sections:In Section II, 
the TDHFB equations are given together with the initial conditions 
of the two colliding superfluid nuclei. 
In Section III, the TDHFB calculations of the collision processes of the 
two oxygens ${}^{20}$O's are carried out with the three values 
of the boost energies, $E_{\rm boost} =$ 4.8, 5.0, and 5.2 MeV, 
leading to the center-of-mass energy $E_{\rm cm}=$ 9.21, 9.41,
and 9.61 MeV, respectively.     
The trajectories, pairing energies, and the number of the transferred nucleons 
are displayed. In Section IV, the effects of the initial relative phase (gauge angle) 
are discussed. Section V is for the summary and concluding remarks.  

\section{Basic equations and initial conditions}
\label{basic_equation}
\subsection{Basic equation}
The nuclear Hamiltonian under consideration is in the form,
\begin{eqnarray}
  H = \sum_{\alpha \beta} T_{\alpha \beta} C_{\alpha}^{\dagger} C_{\beta} 
     + \frac{1}{4} \sum_{\alpha \beta \gamma \delta} 
         {\cal V}_{\alpha \beta \gamma \delta} C_{\alpha}^{\dagger} C_{\beta}^{\dagger}  
                     C_{\delta} C_{\gamma}, \label{origH} 
\end{eqnarray}
where $T_{\alpha \beta}$ is the kinetic energy matrix element and 
${\cal V}_{\alpha \beta \gamma \delta}$ is the antisymmetrized two-body matrix element 
of the Gogny interaction. 
The operator $C_{\alpha}^{\dagger} (C_{\alpha})$ is a nucleon creation (annihilation) 
operator of a state labelled with $\alpha$. 

The quasi-particles ${\beta_{k}^{(\tau)}}^{\dagger}$ and $\beta_{k}^{(\tau)}$ 
are introduced by the Bogoliubov transformation from the particle operators 
$C_{\alpha}^{\dagger}$  and $C_{\alpha}$,  
\begin{eqnarray}
   {\beta_{k}^{(\tau)}}^{\dagger} &=& 
    \sum_{\alpha} \left( U_{\alpha k}^{(\tau)} C_{\alpha}^{\dagger}  
                       + V_{\alpha k}^{(\tau)} C_{\alpha} \right), \label{Bog-1} \\
   \beta_{k}^{(\tau)}           &=& 
    \sum_{\alpha} \left( {U_{\alpha k}^{(\tau)}}^{*} C_{\alpha}
                       + {V_{\alpha k}^{(\tau)}}^{*} C_{\alpha}^{\dagger}   \right)\ ,  \label{Bog-2}
\end{eqnarray}   
where $\tau = $ p (n) for protons (neutrons), respectively.

In the TDHFB method, the equations of motion 
for the matrices $U^{(\tau)}$ and $V^{(\tau)}$ in the Bogoliubov transformation 
(\ref{Bog-1}) and (\ref{Bog-2}) 
are given in the form~\cite{YH-KN,YH-EPJA},  
\begin{eqnarray}
  i \hbar \frac{\partial}{\partial t} 
    \left(
    \begin{array}{c}
                U^{(\tau)}(t) \cr
                V^{(\tau)}(t)
             \end{array}
             \right)
     = {\cal H}^{(\tau)} \left( 
    \begin{array}{c}
                U^{(\tau)}(t) \cr
                V^{(\tau)}(t)
             \end{array}
             \right), \label{tdhfbeq_UV}
\end{eqnarray}
with the HFB Hamiltonian ${\cal H}^{(\tau)}$,  
\begin{eqnarray}
  {\cal H}^{(\tau)} = \left( 
             \begin{array}{cc}
               h^{(\tau)} & \Delta^{(\tau)} \cr
              - {\Delta^{(\tau)}}^{*} & - {h^{(\tau)}}^{*} 
             \end{array}             
             \right)\, .  \label{hfbH}
\end{eqnarray}
The mean field Hamiltonian $h^{(\tau)}$ and the pairing mean field $\Delta^{(\tau)}$ are 
introduced through the relations \cite{RS}, 
\begin{eqnarray}
  h_{\alpha \beta}^{(\tau)} &=& T_{\alpha \beta} + \Gamma_{\alpha \beta}^{(\tau)},  \label{mean-field_hamil}\\ 
 \Gamma_{\alpha \beta}^{(\tau)} &=& \sum_{\gamma \delta} 
      {\cal V}_{\alpha \gamma \beta \delta} \rho_{\delta \gamma}^{(\tau)}, \quad
 \Delta_{\alpha \beta}^{(\tau)} = \frac{1}{2}\sum {\cal V}_{\alpha \beta \gamma \delta} 
                        \kappa_{\gamma \delta}^{(\tau)},     \label{GamDel}
\end{eqnarray}
where $\rho^{(\tau)}$ and $\kappa^{(\tau)}$ are normal density matrix and pairing tensor, 
\begin{eqnarray}
\rho_{\alpha \beta}^{(\tau)}   = \left( {V^{(\tau)}}^{*} {V^{(\tau)}}^{T} \right)_{\alpha \beta}, \, \quad  
\kappa_{\alpha \beta}^{(\tau)} = \left( {V^{(\tau)}}^{*} {U^{(\tau)}}^{T} \right)_{\alpha \beta}, \label{rhoVV_kapVU} 
\end{eqnarray}
respectively. 
The symbol T in Eq. (\ref{rhoVV_kapVU}) stands for the transpose of a matrix.

As the basis functions labeled with $\alpha, \beta, \cdots$, we make use of the 
spatial grid points, i.e., the Lagrange mesh~\cite{Baye,Matsuse} in the direction 
of the z axis, while the two-dimensional harmonic oscillator eigen functions 
are used in the directions of the x and y axes~\cite{YH-PRC88}.   

The HFB matrices ${U_{\alpha k}^{(\tau)}}^{(0)}$ and ${V_{\alpha k}^{(\tau)}}^{(0)}$ 
of the ground state of a nucleus are obtained by solving 
the HFB equations~\cite{RS}, 
\begin{eqnarray}
   & & \left( \begin{array}{cc}
           h^{(\tau)} - \lambda^{(\tau)}        &     \Delta^{(\tau)}  \\
        - {\Delta^{(\tau)}}^*  &  - {h^{(\tau)}}^{*} + \lambda^{(\tau)}
           \end{array}    \right) 
   \left( \begin{array}{c} 
           U_k^{(\tau)}  \\
           V_k^{(\tau)}   
           \end{array}
   \right)    \nonumber \\
 &=& E_k^{(\tau)}  \left( \begin{array}{c} 
           U_k^{(\tau)}  \\
           V_k^{(\tau)}   
           \end{array}
   \right)  ,   \label{hfb-equation}  
\end{eqnarray} 
with the eigenvalues $E_{k}^{(\tau)}$ and a chemical potential $\lambda^{(\tau)}$.
In Eq. (\ref{hfb-equation}), $U_k^{(\tau)}$ ($V_k^{(\tau)}$) 
denotes the vector 
$U_{\alpha k}^{(\tau)}$ ($V_{\alpha k}^{(\tau)}$) corresponding to the eigenvalue 
$E_{k}^{(\tau)}$ of the quasi-particle with a label $k$, respectively. 

The inclusion of the chemical potential or any real variable $\mu$ 
in the mean-field Hamiltonians $h^{(\tau)} - \mu$ and $- {h^{(\tau)}}^{*} + \mu$ in Eq. (\ref{hfbH}) 
will keep unchanged the evolution of the one-body density matrix $\rho^{(\tau)}$ and two-body
correlation matrix $\kappa^{(\tau)} {\kappa^{(\tau)}}^{*}$.

\subsection{Initial conditions}
\begin{figure}[htb]
\begin{center}
\resizebox{0.4\textwidth}{!}{%
  \includegraphics{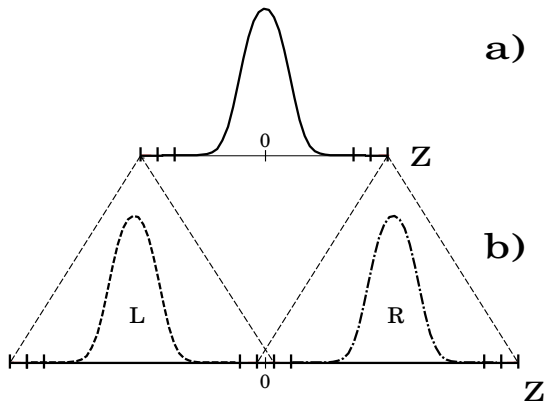}
}
\end{center}
\caption{ 
Initial condition of the TDHFB equation (\ref{tdhfbeq_UV}). 
A HFB ground state is calculated by using the basis functions 
with the number of the grid points $N_{\rm grid}^{(0)} = N_{\rm grid}/2$ (a)).
The HFB ground state in a) is mapped on the space of the 
basis functions with the number of the grid points 
$N_{\rm grid}$ (L or R in b)).
}
\label{initial_position}
\end{figure}

When we set up the initial condition of the TDHFB equation (\ref{tdhfbeq_UV}), 
we assume that the two nuclei are uncorrelated and independent of 
each other if the distance between the two nuclei is large enough.
With the purpose of realizing the situation, we make use of a 
Lagrange mesh whose number of the grid points $N_{{\rm grid}}$ is just the double 
of that of the original Lagrange mesh $N_{{\rm grid}}^{(0)}$ (Fig. \ref{initial_position}): 
$N_{{\rm grid}} = 2 \times N_{{\rm grid}}^{(0)}$.

In the left (L) region with negative z ($z < 0$) in the doubled Lagrange mesh 
(b) in Fig. \ref{initial_position}), 
the HFB matrices ${U_{\alpha k}^{(\tau)}}^{(0)}$ and ${V_{\alpha k}^{(\tau)}}^{(0)}$ 
of the ground state solution on the Lagrange mesh (a) in Fig. \ref{initial_position}) 
are mapped into the matrices $U_{\alpha k}$ and $V_{\alpha k}$ on the doubled 
Lagrange mesh, 
\begin{eqnarray}
   W_{\alpha k}^{(\tau)} = \left \{ 
             \begin{array}{ll}
                {W_{\alpha k}^{(\tau)}}^{(0)}\,\,,   
                                   & \alpha = 1, 2, \cdots, N_{{\rm base}}/2\,, \\
                      0, 
                                   & \alpha = N_{{\rm base}}/2 + 1, \cdots, N_{{\rm base}}\,,
             \end{array}
                           \right. \label{initial-L}
\end{eqnarray} 
where $k = 1, 2, \cdots, N_{{\rm base}}/2$ and $W^{(\tau)} ({W^{(\tau)}}^{(0)})$ 
stands for the matrix $U^{(\tau)} ({U^{(\tau)}}^{(0)})$ or $V^{(\tau)} ({V^{(\tau)}}^{(0)}) $, respectively. 
Here, $N_{\rm base}$ is the total number of the basis functions of the 
two-dimensional harmonic oscillator eigenfunctions and Lagrange mesh 
together with the spin degrees of freedom.

Just in the same way, in the right (R) region with positive z ($0 < z$) 
in the doubled Lagrange mesh, we have the mapped matrices $U_{\alpha k}^{(\tau)}$ 
and $V_{\alpha k}^{(\tau)}$,
\begin{widetext}
\begin{eqnarray}
   W_{\alpha k}^{(\tau)} = \left \{ 
             \begin{array}{ll}
                      0,            & \alpha = 1, 2, \cdots, N_{{\rm base}}/2\,, \\
                {W_{\alpha^{\prime} k}^{(\tau)}}^{(0)}\,\,  (\alpha^{\prime} = \alpha - N_{{\rm base}}/2)\,,   
                                    & \alpha = N_{{\rm base}}/2 + 1, \cdots, N_{{\rm base}}\,, 
             \end{array}
                           \right.  \label{initial-R}
\end{eqnarray} 
\end{widetext}
where $k = N_{{\rm base}}/2 + 1, \cdots, N_{{\rm base}}$ and 
$W^{(\tau)} ({W^{(\tau)}}^{(0)})$ 
is used for the matrix $U^{(\tau)} ({U^{(\tau)}}^{(0)})$ or $V^{(\tau)} ({V^{(\tau)}}^{(0)}) $, respectively.

Note that this initialization method conserves the fermion commutation relations \cite{RS},
\begin{eqnarray}
{U^{(\tau)}}^{\dagger} U^{(\tau)} + {V^{(\tau)}}^{\dagger} V^{(\tau)} = 1, \label{unitarity-1} \\
U^{(\tau)}{U^{(\tau)}}^{\dagger} + {V^{(\tau)}}^{*} {V^{(\tau)}}^{T} = 1, \\
{U^{(\tau)}}^{T} V^{(\tau)} + {V^{(\tau)}}^{T} U^{(\tau)} = 0,\\
U^{(\tau)} {V^{(\tau)}}^{\dagger} + {V^{(\tau)}}^{*} {U^{(\tau)}}^{T} = 0.  \label{unitarity-4}
\end{eqnarray}

Each of the two nuclei is boosted with a momentum so that 
the total momentum of the system is zero and the initial position of the 
center-of-mass is kept at the initial point.  

\section{Head-on collisions of two ${}^{20}$O's}
In the present calculations, the Gogny D1S is used as the effective interaction. 
We note that the Coulomb force is used only in the mean-field part 
and is not included in the pairing part of the HFB Hamiltonian.

The parameters used in the calculations are as follows:
The grid spacing $\Delta z$ = 0.91 fm 
and the total number of the grid points of the doubled Lagrange mesh $N_{\rm grid}$ 
is forty six. The harmonic oscillator eigenfunctions are used 
as the basis functions in the x-y plane. 
The space of the harmonic oscillator quantum number is restricted 
as $n_x + n_y \leq {N_{\rm shell}} = 4$ with the quantum number $n_x$ ($n_y$) in the 
direction of the x (y) axis, respectively. 
The total number of the basis functions $N_{\rm base}$ is 
$N_{\rm base} = (N_{\rm shell} + 1) (N_{\rm shell} + 2) N_{\rm grid}$ 
including spin up and down.
The harmonic oscillator parameter $\hbar \omega = 14.6$ MeV 
and the maximum number $N_{\rm shell}$ are used in the calculations of the 
HFB ground state solutions (a) in Fig. \ref{initial_position}) 
as well as the head-on collisions of the two nuclei  
on the doubled mesh space (b) in Fig. \ref{initial_position}) . 

Here, we note that the restricted space limits the number 
of the degrees of freedom to 2760 including spin and isospin 
while a full calculation in Cartesian mesh would involve around $10^5$ degrees of freedom.
The CPU time for the one step of the integration of the TDHFB equations is 
four minutes on HITACHI SR16000M1 by using 512 CPUs. 
The one trajectory in the subsequent calculations is carried out in 
eight to ten days.

\begin{figure}[htb]
\begin{center}
\resizebox{0.4\textwidth}{!}{%
  \includegraphics{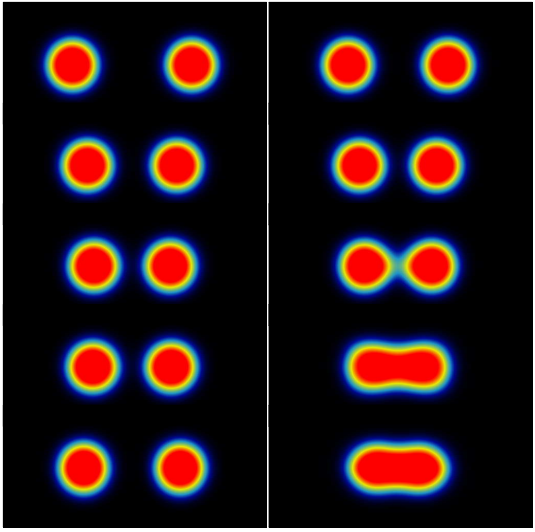}
}
\end{center}
\caption{ Densities in the yz plane in the collisions ${}^{20}$O + ${}^{20}$O 
for the center-of-mass energies $E_{\rm cm}$=9.21 MeV (left) 
and $E_{\rm cm}$=9.61 MeV (right). 
The densities are shown with respect to the time 
at the times $c t$ = 180, 360, 540, 720, and 900 fm 
from the top to the bottom, respectively. $c$ is the light speed.
}
\label{film}
\end{figure}

\subsection{Relative distance, relative momentum, and transferred number of particle}
The evolution of the system is shown in Fig. \ref{film} 
for the two center-of-mass energies $E_{\rm cm}$, 
where one is below the barrier and another is above the barrier. 
For the energy $E_{\rm cm}$ below the barrier top, 
the two nuclei exchange nucleons 
and separate into two fragments, while the two nuclei merge into one  
for the energy $E_{\rm cm}$ above the barrier top. 

 In order to understand the dynamical properties 
of the collision processes around the barrier, 
we followed Washiyama's method of introducing a section plane at a point 
on the z axis between the two colliding nuclei~\cite{washiyama_PRC78}.
The section plane is set at the point $z = z_{\rm s}$ where the density 
$\tilde{\rho}_{{\rm L}}({\bf r}, t)$ is equal to the density 
$\tilde{\rho}_{{\rm R}}({\bf r}, t)$.  
Here, $\tilde{\rho}_{{\rm L/R}}({\bf r}, t)$ is the density which 
started from the nucleus in the left (L) (right (R)) region,   
\begin{eqnarray}
 \tilde{\rho}_{{\rm L/R}} ({\bf r}, t) 
   = \sum_{\alpha \beta} \Phi_{\alpha}({\bf r}) \Phi_{\beta}^{*}({\bf r}) 
     \tilde{\rho}_{\alpha \beta}^{{\rm L/R}} \delta_{\sigma_{\alpha} \sigma_{\beta}},                         \label{densityLR}
\end{eqnarray}   
with the density matrix $\tilde{\rho}_{\alpha \beta}^{{\rm L/R}}$, 
\begin{eqnarray}
  \tilde{\rho}_{\alpha \beta}^{{\rm L/R}} 
        = \sum_{\tau\, =\, {\rm p}, {\rm n} } \sum_{k^{\prime}} 
           {V_{\alpha k^{\prime}}^{(\tau)}}^{*} V_{\beta k^{\prime}}^{(\tau)},  \label{dmatrixLR}
\end{eqnarray}
respectively. 
The index $k^{\prime}$ takes the values $k^{\prime} = 1, 2, \cdots, N_{{\rm base}}/2$ 
for the $\tilde{\rho}_{\alpha \beta}^{{\rm L}}$ and 
$k^{\prime} = N_{{\rm base}}/2 +1, N_{{\rm base}}/2 + 2, \cdots, N_{{\rm base}}$ 
for the $\tilde{\rho}_{\alpha \beta}^{{\rm R}}$.
In (\ref{densityLR}), $\Phi_{\alpha}({\bf r})$ and $\Phi_{\beta}({\bf r})$ 
are the basis functions of the two-dimensional harmonic oscillator eigen functions 
and Lagrange mesh~\cite{YH-PRC88}, 
and $\sigma_{\alpha}$ and so on are the labels of the spin.

Making use of the section plane at $z = z_{\rm s}$, we calculated the 
number of the nucleons in each of the left region and right region 
with respect to the plane, 
\begin{eqnarray}
  m_{\rm L} &=& \int d^3 x \,\, \rho({\bf r}) \theta(z_{\rm s} - z)\, ,  \label{eq:m_L} \\
  m_{\rm R} &=& \int d^3 x \,\, \rho({\bf r}) \theta(z - z_{\rm s})\, ,
\end{eqnarray}  
with the total density $\rho({\bf r}) = \rho^{(\rm p)}({\bf r}) + \rho^{(\rm n)}({\bf r})$ 
made from the proton (p) and neutron (n) densities. 

The center-of-mass position $z_{\rm L}$ $(z_{\rm R})$ 
and momentum $p_{\rm L}$ $(p_{\rm R})$ in the left (right) region divided by 
the section plane are calculated, 
\begin{eqnarray}
  z_{\rm L} = {\rm Tr} \left[ (z)^{(\rm L)} \rho \right] / m_{\rm L}, \quad 
  z_{\rm R} = {\rm Tr} \left[  (z)^{(\rm R)} \rho \right] / m_{\rm R} ,   \label{comLR}
\end{eqnarray}
and
\begin{eqnarray}
  p_{\rm L} = {\rm Tr} \left[  (p_{z})^{(\rm L)} \rho \right], \quad 
  p_{\rm R} = {\rm Tr} \left[  (p_{z})^{(\rm R)} \rho \right] ,           \label{pzLR}
\end{eqnarray}  
respectively.
In Eqs. (\ref{comLR}) and (\ref{pzLR}) , the notation Tr  is the trace of a matrix,  
and the notations $(z)^{({\rm L})}$, $(z)^{({\rm R})}$, 
$(p_{z})^{({\rm L})}$ and $(p_{z})^{({\rm R})}$ stand for the matrices 
with the matrix elements, 
\begin{widetext}
\begin{eqnarray}
  (z)^{({\rm L})}_{\alpha \beta} 
 &=& \int d^3 x\,\, \theta (z_{\rm s} - z) \Phi_{\alpha}({\bf r})^{*} z  \Phi_{\beta}({\bf r})  
                            \delta_{\sigma_{\alpha} \sigma_{\beta}}\, , \quad 
  (p_z)^{({\rm L})}_{\alpha \beta} 
 = \int d^3 x\,\, \theta (z_{\rm s} - z) \Phi_{\alpha}({\bf r})^{*} 
         \left(- i \hbar \frac{\partial}{\partial z} \right )  \Phi_{\beta}({\bf r})  
                            \delta_{\sigma_{\alpha} \sigma_{\beta}}\, ,  \\
  (z)^{({\rm R})}_{\alpha \beta} 
 &=& \int d^3 x\,\, \theta (z - z_{\rm s}) \Phi_{\alpha}({\bf r})^{*} z  \Phi_{\beta}({\bf r}) 
                            \delta_{\sigma_{\alpha} \sigma_{\beta}}\, ,    \quad      
  (p_z)^{({\rm R})}_{\alpha \beta} 
 = \int d^3 x\,\, \theta (z - z_{\rm s}) \Phi_{\alpha}({\bf r})^{*} 
         \left(- i \hbar \frac{\partial}{\partial z} \right ) \Phi_{\beta}({\bf r}) 
                            \delta_{\sigma_{\alpha} \sigma_{\beta}}.
\end{eqnarray}
\end{widetext}
Here, $\rho$ is the total density matrix $\rho = \rho^{(p)} + \rho^{(n)}$.
The relative coordinate $R$ and relative momentum $P_{z}$ are defined, 
\begin{eqnarray}
   R &=& z_{\rm R} - z_{\rm L}\,, \nonumber \\ 
 P_z &=& \left( m_{\rm L} p_{\rm R} - m_{\rm R} p_{\rm L} \right) 
                                         / ( m_{\rm L} + m_{\rm R} )\, .  
\end{eqnarray}

\subsection{Trajectories, pairing energies, and number of the transferred particles}
As the examples of the head-on collisions of the two ${}^{20}$O's, 
we chose three boost energies $E_{\rm boost} = 4.8$ MeV, 5.0 MeV, 
and 5.2 MeV, leading to the center-of-mass energies $E_{\rm cm} =$ 9.21, 9.41, 
and 9.61 MeV, respectively.

\begin{figure}[htb]
\begin{center}
\resizebox{0.4\textwidth}{!}{%
  \includegraphics{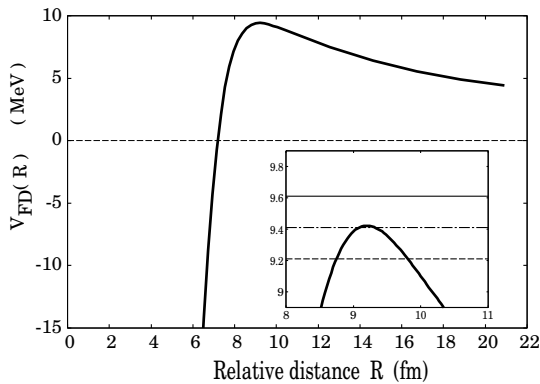}
}
\end{center}
\caption{The frozen density potential $V_{\rm FD}(R)$ 
with respect to relative coordinate $R$ 
of the head-on collision of two ${}^{20}$O's. 
The inset figure is the magnification of the region around the top of the 
frozen density potential. In the inset figure, three energies 
are described with broken, chain, and solid lines for 
$E_{\rm cm}$ = 9.21, 9.41, and 9.61 MeV, respectively.  
}
\label{fdpotential}
\end{figure}

In Fig. \ref{fdpotential}, we display the frozen density potential $V_{\rm FD}(R)$ 
with respect to the relative distance $R$ together with the 
positions of the three initial energies to show that they are 
below, nearly on, and above the top of the frozen density potential, respectively.

Corresponding to the three energies $E_{\rm cm}$, we got three trajectories 
of the colliding oxygens ${}^{20}$O's in the phase space $R - P_{z}$ 
of the relative distance $R$ and the relative momentum $P_{z}$.

\begin{figure}[htb]
\begin{center}
\resizebox{0.4\textwidth}{!}{%
  \includegraphics{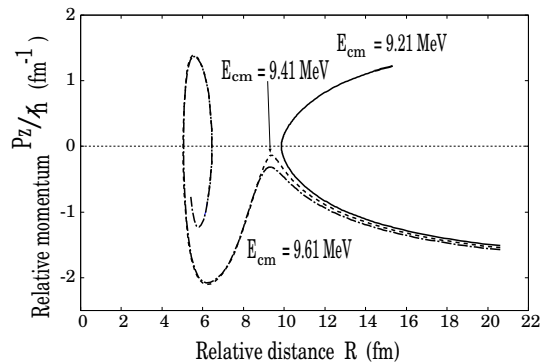}
}
\end{center}
\caption{Trajectories in the phase space of the relative coordinate $R$ and 
the relative momentum $P_z$. The curve in solid (broken, chain) corresponds to the 
initial energy $E_{\rm cm} =$ 9.21 (9.41, 9.61) MeV, respectively. 
The two trajectories with $E_{\rm cm}$ = 9.41 and 9.61 MeV almost fully overlap 
in the region $R <$ 8.5 fm. 
}
\label{trajectories}
\end{figure}

In Fig. \ref{trajectories}, we display the three trajectories in the phase space 
$R - P_{z}$. 
They start at $R = 20.91$ fm and follow each other side by side 
till they come around $R \sim$ 12 fm. 
After passing the point around $R \sim$ 12 fm, the three trajectories 
begin to separate from each other: 
The trajectory with the energy $E_{\rm cm}$ = 9.21 MeV corresponds to  
a process in which the two oxygens ${}^{20}$O's come near each other,  
stop at the turning point around $R \sim$ 10 fm, 
and then bounce back into the two 
fragments that are mixture of the components 
including the transfer states, the two initial ${}^{20}$O's states 
in their ground states, and in their excited states.

The two trajectories with the energies $E_{\rm cm}$ = 9.41 MeV and 9.61 MeV, 
on the other hand, represent the process of fusion of the two nuclei after 
slowing down in the relative motion.
The combined systems display vibration after the fusion of the two ${}^{20}$O's 
in both cases of these energies. 

\begin{figure}[htb]
\begin{center}
\resizebox{0.4\textwidth}{!}{%
  \includegraphics{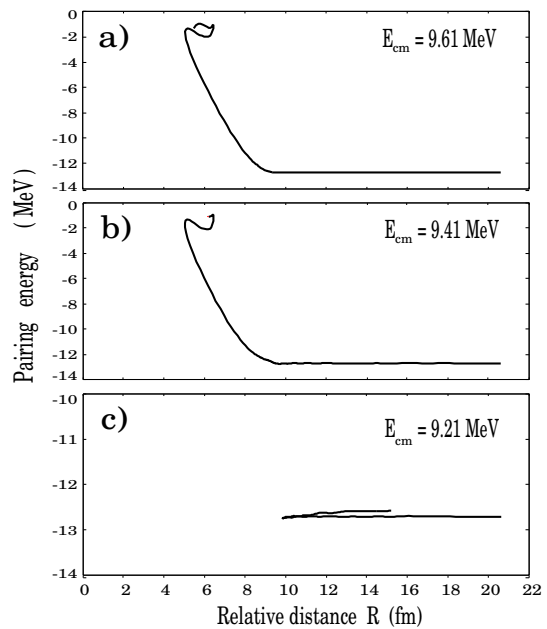}
}
\end{center}
\caption{Pairing energies of the trajectories with the energies 
$E_{\rm cm}$ = 9.61 MeV (a)), 9.41 MeV (b)), and 9.21 MeV (c)) are plotted 
with respect to the relative distance $R$. } 
\label{epgss_R}
\end{figure}

In Fig. \ref{epgss_R}, the variations of the pairing energies along the three 
trajectories in Fig. \ref{trajectories} are displayed. 
We see that the pairing energies are kept almost constant 
before the two nuclei come to the region of the top of the potential energy 
$V_{\rm FD}(R)$ at around $R \sim$ 9.2 fm for the cases of the energies 
$E_{\rm cm} =$ 9.41 MeV and 9.61 MeV. 
In a similar way, the pairing energy is kept almost constant until 
the two nuclei come to the turning point 
at around $R \sim$ 10.0 fm for the case 
with the energy $E_{\rm cm}$ = 9.21 MeV.
   
Once the two nuclei begin to fuse in the cases of the energies 
$E_{\rm cm}$ = 9.41 and 9.61 MeV, the pairing energy $E_{\rm pair}$ 
rapidly becomes small in the magnitude from -13 MeV to -2 MeV, 
and oscillates around the value $E_{\rm pair} \sim$ -2 MeV in each case 
of the energies (a) and b) in Fig. \ref{epgss_R}). 
We can interpret this diminution of the pairing energy 
from the increase of the excitation energy after the fusion.
The occurrence of the internal excitations 
is expected to reduce the pairing correlation. 
When the two nuclei turn back from the turning point (c) in Fig. \ref{epgss_R}), 
on the other hand, the variation of the pairing energy is 0.15 MeV.

In relation with the variations of the pairing energies in Fig. \ref{epgss_R}, 
let us calculate the numbers of the transferred nucleons.
We followed the Washiyama${}^{\\,}$s definition of the number 
of the transferred nucleons $N_{\rm trans}^{{\rm L}/{\rm R}} \left[R(t)\right]$ 
in the collision process 
of a pair of the identical nuclei, which was used 
in the framework of the TDHF~\cite{washiyama_PRC78}, 
\begin{eqnarray}
  N_{\rm trans}^{{\rm L}} \left[R(t)\right] 
     = \int d^3x {\tilde{\rho}}_{{\rm L}} ({\bf r}, t) \theta(z), 
\end{eqnarray}
where L stands for the left nuclei at the initial time $t =$ 0   
and the density ${\tilde{\rho}}_{{\rm L}/{\rm R}} ({\bf r}, t)$ is 
given in (\ref{dmatrixLR}). We paid attention only to the transfer of the 
nucleons which were in the left nuclei at the initial time, 
since the present system of two oxygen nuclei ${}^{20}$O's is 
symmetric with respect to the origin $z =$ 0. 
Then, the section plane is also put at the origin $z_{s} =$ 0. 
 
\begin{figure}[htb]
\begin{center}
\resizebox{0.4\textwidth}{!}{%
  \includegraphics{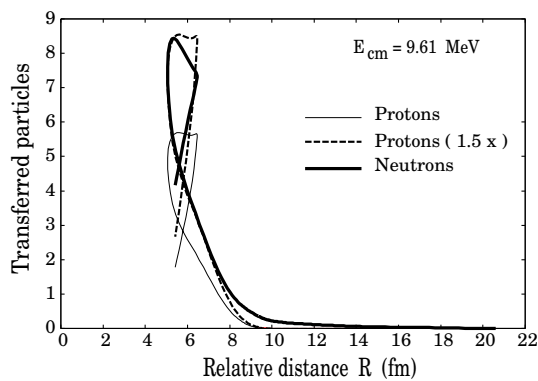}
}
\end{center}
\caption{ 
Transferred numbers of the protons (thin curve) and neutrons (thick curve) 
with respect to the relative distance along the trajectory with the 
energy $E_{\rm cm}$ = 9.61 MeV. The broken curve is plotted by 
multiplying the number of the transferred protons by a factor of 1.5.
}
\label{transfer-hi}
\end{figure}

\begin{figure}[htb]
\begin{center}
\resizebox{0.4\textwidth}{!}{%
 \includegraphics{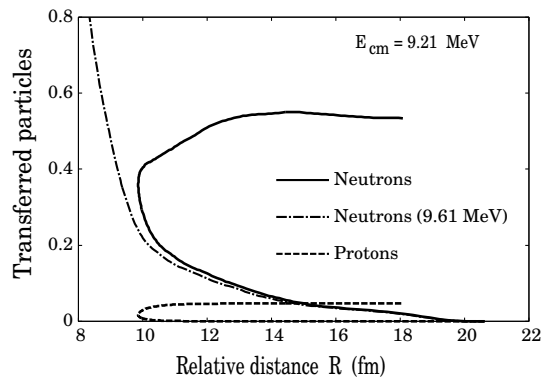}
}
\end{center}
\caption{ 
Transferred numbers of the protons (broken curve) and neutrons (solid curve) 
with respect to the relative distance in the trajectory with the 
energy $E_{\rm cm}$ = 9.21 MeV. 
The chain curve is for the transferred number 
of the neutrons in Fig. \ref{transfer-hi} as a reference. 
}
\label{transfer-lo}
\end{figure}

In Figs. \ref{transfer-hi} and \ref{transfer-lo}, 
we show the transferred numbers of the protons and neutrons 
with respect to the relative distance $R$ 
in the cases of the trajectories with the energies 
$E_{\rm cm}$ = 9.61 MeV and 9.21 MeV, respectively.
 
In Fig. \ref{transfer-hi}, the two nuclei are in the process of the fusion 
within the region $R \leq$ 9.2 fm, and the transferred numbers of the protons 
and neutrons rapidly increase as the overlap of the two nuclei becomes larger.
In the figure, multiplying the transferred number of the protons by 1.5, 
we get a curve which goes along the curve of the number 
of the transferred neutrons.
Since the value 1.5 is just the number of the N/Z ratio in the ${}^{20}$O,  
we see that the protons and neutrons begin to move into the region 
of the other nuclei in a similar way after the combined system passes over 
the top of the potential energy $V_{\rm FD}(R)$.  

In the region $R \geq$ 9.2 fm, the number of the transferred protons is 
practically zero, while the number of the transferred neutrons  
slowly increases to the value of around 0.4 
as the two nuclei approach each other.
  
In Fig. \ref{transfer-lo}, the energy of the system  
is below the top energy of the potential curve $V_{\rm FD}(R)$.  
The numbers of the transferred nucleons are small along the trajectory 
in which the system approaches each other, stop at the turning point, 
and separate again into two fragments.  

The number of the transferred protons increase only when the system is near 
the turning point, though it is much smaller than that of the neutrons 
under the influence of the Coulomb potential.
 
The number of the transferred neutrons increases monotonously up to 
0.2 as the two nuclei approach each other before they come to the turning point 
$R \sim$ 10.0 fm. The number of the transferred neutrons jumps up 
by 0.2 near the turning point, and comes up to 0.55 at the end point 
of the curve in Fig. \ref{transfer-lo} at the distance $R \sim$ 18 fm.   
As a result, the transferred number of the neutrons is more than ten times 
as large as that of the protons in the whole process in Fig. \ref{transfer-lo}.
The ratio is much larger than the N/Z ratio 1.5 in the ${}^{20}$O.
The Coulomb barrier is the main reason for the difference 
between the transferred numbers of the protons' and that of the neutrons'.

In Fig. \ref{transfer-lo}, we also note that the curve 
of the transferred neutrons of the trajectory 
with the energy $E_{\rm cm}$ = 9.21 MeV follows that of the 
trajectory with the energy $E_{\rm cm}$ = 9.61 MeV 
in Fig. \ref{transfer-hi} in the approaching stage of the two nuclei 
before they come to the region near the turning point.
This could be understood from the variations of the pairing energies 
in Fig. \ref{epgss_R}: We see that the pairing energies of the 
trajectories of the three cases of the energies 
are kept practically constant at the initial values 
during the two nuclei approach each other.  
The tail parts of the neutron densities of the approaching nuclei ${}^{20}$O's  
with the same pairing energies are almost the same to each other.  
Then, the transferred numbers of the neutrons 
during the approaching stage of the two nuclei could be in the common tendency 
among the three cases of the energies. 

\section{Effects of the initial relative gauge angle}
In the superfluid nuclei which satisfy the HFB equations (\ref{hfb-equation}), 
it is well known that there is a gauge invariance in relation with the 
transformation in terms of an operator $G(\chi) = e^{- i \chi \hat{N}}$ 
with the number operator $\hat{N}$.
The operator $G(\chi)$ transforms the matrices 
$U_{\alpha k}$ and $V_{\alpha k}$ to $e^{-i \chi} U_{\alpha k}$ 
and $e^{i \chi} V_{\alpha k}$, respectively. 
The transformation does not change the properties of the nucleus 
in the ground state as long as an isolated nucleus is under consideration.
In the collision process of the two superfluid nuclei, however, 
we do not know in advance the effects of the gauge transformations 
in the two nuclei at the initial time on the properties 
of the colliding two nuclei on the trajectory.  

From the quantum mechanical point of view, 
the degree of of freedom of the 
transformation by the operator $G(\chi)$ is made use of 
to project out a state with a specified number of particles 
from the HFB state.

In the framework of the mean-field of the (TD)HFB, 
what we would like to study is 
the influence of the gauge transformations 
within the HFB ground states at the initial stage of the colliding nuclei   
on the behavior of the system later in the collision process.

\subsection{A combination of the ${}^{16}$O and ${}^{20}$O}
As an example of the case in which the gauge transformation of the 
superfluid nucleus plays no effects on the collision process, 
we take the combination of the ${}^{16}$O and ${}^{20}$O.
We adopt the phase factors $e^{i \chi}$ with $\chi = 0, 45, 90,$ and 135 degrees 
as the representation of the gauge transformation 
in the superfluid nuclei ${}^{20}$O, 
\begin{eqnarray}
  U_{\alpha k} = e^{i \chi}   U_{\alpha k}^{(0)}, \,\,  \quad 
  V_{\alpha k} = e^{- i \chi} V_{\alpha k}^{(0)}.  \label{o16-o20-angle}
\end{eqnarray} 
The ${}^{16}$O is not in the superfluid phase but in the normal state 
on both sides of the protons and neutrons. 
Note that this transformation keeps unchanged the normal density, 
and modify the pairing tensor $\kappa$ with a phase shift $2 \chi$. 
In consequence the values of $\chi$ between 180 and 360 degrees are redundant 
with the values between 0 and 180 degrees.

\begin{figure}[htb]
\begin{center}
\resizebox{0.4\textwidth}{!}{%
  \includegraphics{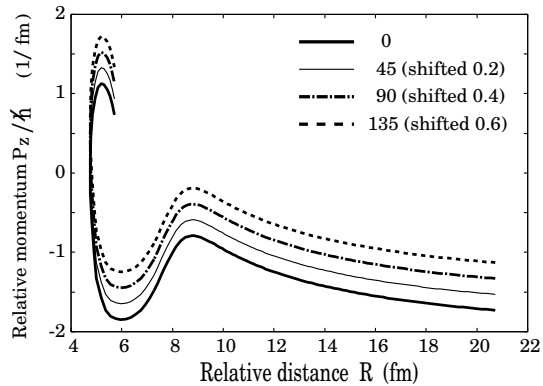}
  
}
\end{center}
\caption{
Trajectories of the head-on collisions of the ${}^{16}$O and 
the ${}^{20}$O 
with the energy $E_{\rm cm} =$ 11.41 MeV  
in the phase space of the relative distance $R$ 
and relative momentum $P_z$. 
The solid (thin solid, chain, and broken) curve 
is for the case with the initial gauge angle $\chi$ = 0 
(45, 90, and 135) degree(s), respectively. 
All of the trajectories overlap each other. 
The curve with $\chi$ = 45 (90, 135) degrees is shifted upward 
by 0.2 (0.4, 0.6) $1/{\rm fm}$ for the ease of the eyes, respectively.  
}
\label{ox16-ox20-rz-pz-7.0}
\end{figure}

\begin{figure}[htb]
\begin{center}
\resizebox{0.4\textwidth}{!}{%
  \includegraphics{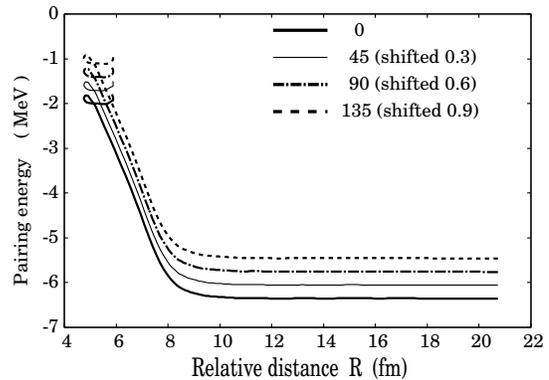}
}
\end{center}
\caption{ 
Variations of the pairing energies with respect to the relative distance 
in the collision processes in Fig. \ref{ox16-ox20-rz-pz-7.0}.
The solid (thin solid, chain, and broken) curve 
is for the case with the initial gauge angle $\chi$ = 0 
(45, 90, and 135) degree(s), respectively. 
All of the curves overlap each other. 
The curve with $\chi$ = 45 (90, 135) degrees is shifted upward 
by 0.3 (0.6, 0.9) MeV for the ease of the eyes, respectively. 
}
\label{ox16-ox20-epgss-7.0}
\end{figure}

We put the ${}^{16}$O as the left-hand side nucleus 
and the ${}^{20}$O with the phase factors in Eq. (\ref{o16-o20-angle}) 
as the right-hand side nucleus in the initial conditions 
(\ref{initial-L}) and (\ref{initial-R}). 
In the case of the energy $E_{\rm cm}$ = 11.41 MeV, 
the trajectories are shown in Fig. \ref{ox16-ox20-rz-pz-7.0}, 
with the different initial gauge angles $\chi$ = 0, 45, 90, and 135 degrees.  
All of the trajectories in Fig. \ref{ox16-ox20-rz-pz-7.0} 
with different initial gauge angles overlap each other completely. 

The variations of the pairing energies with respect to the relative distance 
are shown in Fig. \ref{ox16-ox20-epgss-7.0} in the cases of the different 
initial gauge angles $\chi$. 
After passing the region of the top of the potential energy 
at $R \sim$ 8.5 fm the pairing energies smoothly decrease 
to - 2 MeV and oscillate about the value.
Again, the curves of the pairing energies with different initial gauge angles 
overlap each other completely. 

Thus, we see that there is no dependence on the initial gauge angles 
in the collision process of the combination of a nucleus in the normal state 
and a superfluid nucleus. 
 This result can be understood simply from the fact 
that the normal nucleus does not break the gauge angle symmetry. 
In consequence, changing the phase of the superfluid nuclei is equivalent to 
changing the whole phase of the system, 
in which transformation the evolution of the observables 
is kept unchanged. 

\subsection{A combination of the ${}^{20}$O and ${}^{20}$O}

\begin{figure}[htb]
\begin{center}
\resizebox{0.4\textwidth}{!}{%
  \includegraphics{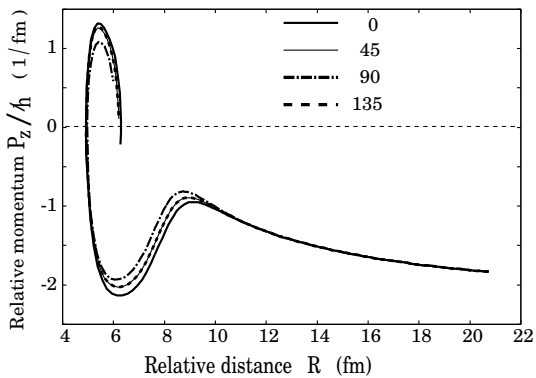}
}
\end{center}
\caption{
Trajectories of the head-on collisions of the ${}^{20}$O and 
the ${}^{20}$O with the energy $E_{\rm cm} =$ 11.41 MeV 
in the phase space of the relative distance $R$ 
and relative momentum $P_z$. 
Each of the trajectories is the result of the first 2000 steps 
of the time integration of the TDHFB equations 
(\ref{tdhfbeq_UV}).  
The solid (thin solid, chain, and broken) curve 
is for the case with the initial gauge angle $\chi$ = 0 
(45, 90, and 135) degree(s), respectively. 
The trajectories with the initial gauge angles $\chi$ = 45 and 135 
degrees overlap each other.
}
\label{rz-pz-7.0-1}
\end{figure}

\begin{figure}[htb]
\begin{center}
\resizebox{0.4\textwidth}{!}{%
  \includegraphics{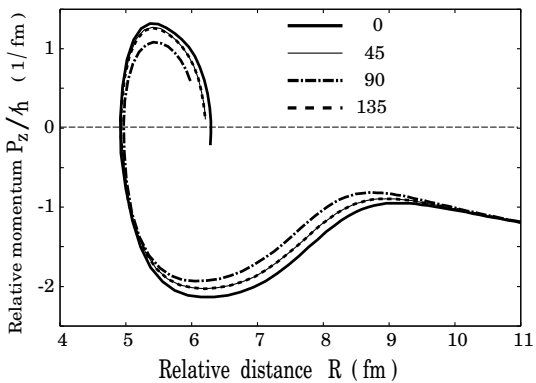}
}
\end{center}
\caption{
Magnification of the region of the relative distance $R \leq$ 11 fm 
in Fig. \ref{rz-pz-7.0-1}.  
}
\label{rz-pz-7.0-2}
\end{figure}

In the case of the combination of the ${}^{20}$O and ${}^{20}$O, 
we follow the way of setting the initial condition of the TDHFB equations 
which is stated in the previous subsection for the case of the 
collisions of the ${}^{16}$O and the ${}^{20}$O. 
The initial phase factors $e^{i \chi}$ with $\chi = 0, 45, 90,$ and 135 degrees 
are multiplied on the HFB solutions $U_{\alpha k}^{(0)}$ and $V_{\alpha k}^{(0)}$ 
of the ground state of the ${}^{20}$O just as in (\ref{o16-o20-angle}). 
Then each of the set of the matrices 
$e^{i \chi} U_{\alpha k}^{(0)}$ and $e^{-i \chi}V_{\alpha k}^{(0)}$ 
are mapped as the right-hand side nucleus, 
while the HFB solutions $U_{\alpha k}^{(0)}$ and $V_{\alpha k}^{(0)}$, 
which is just the case with the gauge angle $\chi = 0$,  
are mapped as the left-hand side nucleus.  

\begin{figure}[htb]
\begin{center}
\resizebox{0.4\textwidth}{!}{%
  \includegraphics{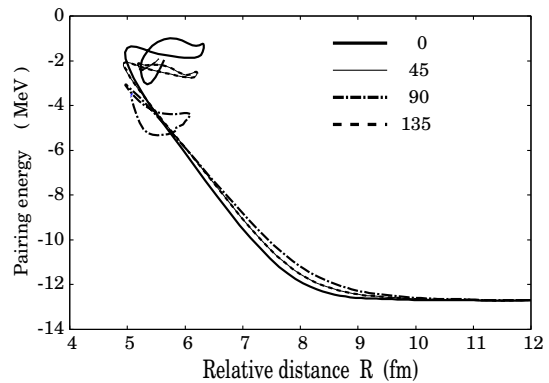}
}
\end{center}
\caption{
Variation of the pairing energies with respect to the 
relative distance $R$ in the trajectories 
in Fig. \ref{rz-pz-7.0-2}.
}
\label{epgss-rz-7.0}
\end{figure}
 
In Figs. \ref{rz-pz-7.0-1} and \ref{rz-pz-7.0-2}, 
we show the trajectories in the case of the energy $E_{\rm cm}$ = 11.41 MeV. 
The latter figure Fig. \ref{rz-pz-7.0-2} is a magnification of the region 
with the relative distance 4 fm $\leq R \leq$ 11 fm 
in the former figure Fig. \ref{rz-pz-7.0-1}. 

In contrast to the cases of the combination of the ${}^{16}$O 
and the ${}^{20}$O in the previous subsection, 
the trajectories are dependent on their initial gauge angles  
and are separated from each other in the region of the relative distance 
$R \leq$ 10 fm. The shift of the trajectory with the initial gauge angle 
$\chi$ = 90 degrees from the one with the angle $\chi$ = 0 degree is the 
largest. The trajectories with the angles $\chi$ = 45 and 135 degrees 
overlap each other and come in the space between the trajectories 
with the angles $\chi$ = 0 and 90 degree(s).   

Just as in the cases of the trajectories, 
the variation of the pairing energy with respect to the relative distance 
$R$ changes as the gauge angle $\chi$ is varied.
In Fig. \ref{epgss-rz-7.0}, we plot the pairing energies of the trajectories 
in Fig. \ref{rz-pz-7.0-2} with respect to the 
relative distance $R$ with the initial gauge angles $\chi = 0, 45, 90$, and $135$ 
degrees. The size of the shift of the pairing energy with the gauge angle 
$\chi$ = 90 degrees from that with $\chi$ = 0 degree is the largest 
among the three cases of the gauge angles.

\begin{figure}[htb]
\begin{center}
\resizebox{0.4\textwidth}{!}{%
  \includegraphics{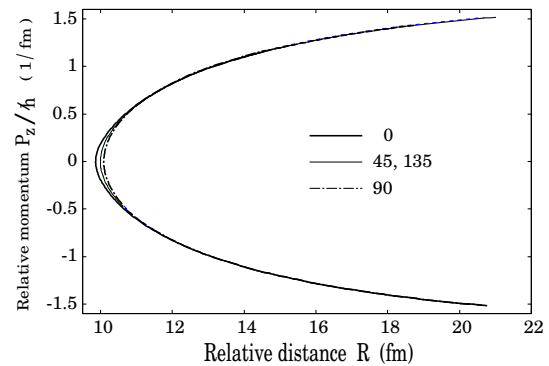}
}
\end{center}
\caption{ 
Trajectories of the head-on collisions of the ${}^{20}$O and 
the ${}^{20}$O with the energy $E_{\rm cm}$ = 9.21 MeV 
in the phase space of the relative distance $R$ 
and relative momentum $P_z$. 
The solid (chain) curve 
is for the case with the initial gauge angle $\chi$ = 0 
(90) degree(s), respectively. 
The thin solid curve is for both $\chi$ = 45 and 135 degrees.
The trajectories with the angles $\chi$ = 45 and 135 
degrees overlap each other.
}
\label{rz-pz-4.8-1}
\end{figure}

\begin{figure}[htb]
\begin{center}
\resizebox{0.4\textwidth}{!}{%
  \includegraphics{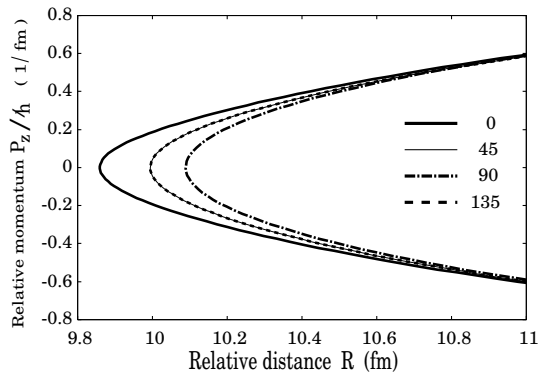}
}
\end{center}
\caption{ 
Magnification of the region of the relative distance 9.8 fm $\leq$ $R \leq$ 11 fm 
in Fig. \ref{rz-pz-4.8-1}. 
The solid (thin solid, chain, and broken) curve is for 
the case with the gauge angle 
$\chi$ = 0 (45, 90, and 135) degree(s), respectively.
}
\label{rz-pz-4.8-2}
\end{figure}

In Figs. \ref{rz-pz-4.8-1} and \ref{rz-pz-4.8-2}, 
we show the trajectories in the case of the energy $E_{\rm cm}$ = 9.21 MeV. 
The latter figure Fig. \ref{rz-pz-4.8-2} is a magnification of the region 
with the relative distance 9.8 fm $\leq R \leq$ 11 fm 
in the former figure Fig. \ref{rz-pz-4.8-1}. 

Just in the same way as the case with the energy $E_{\rm cm} = $ 
11.41 MeV, the trajectories are dependent on the initial gauge angles $\chi$, 
and the shift of the trajectory with the angle $\chi$ = 90 degrees 
from that with the angle $\chi$ = 0 is the largest. 
The two trajectories with the angles $\chi$ = 45 and 90 degrees overlap 
each other.

A remarkably different point in the trajectories with the energy 
$E_{\rm cm}$ = 11.41 MeV and $E_{\rm cm}$ = 9.21 MeV is as follows: 
In the case of the energy $E_{\rm cm}$ = 11.41 MeV, 
the point of the relative distance at which the relative momentum $P_{z}$
of the trajectory takes the smallest absolute value just before 
the fusion shifts to the direction of the small value of the $R$ 
when the initial gauge angle is varied from 0 to 90 degrees through 45 degrees. 
On the other hand, in the case of the energy 
$E_{\rm cm}$ = 9.21 MeV, the turning point of the trajectory 
shifts toward the large value of the $R$ when the gauge angle is 
varied as $\chi =$ 0, 45, and 90 degrees.

\begin{figure}[htb]
\begin{center}
\resizebox{0.4\textwidth}{!}{%
  \includegraphics{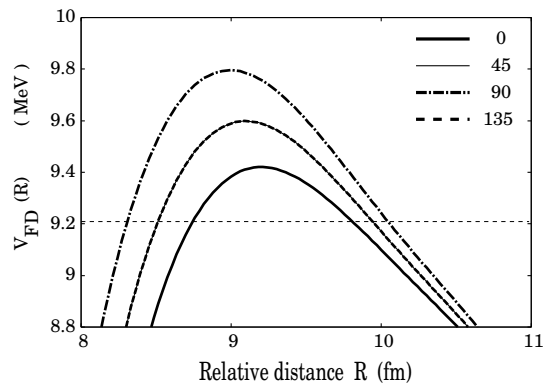}
}
\end{center}
\caption{ 
The frozen density potentials $V_{\rm FD} (R)$ with respect to the 
relative distance $R$ with the initial gauge angle $\chi =$ 0 (solid), 
45 (thin solid), 90 (chain), and 135 (broken) degrees. 
The thin broken line stands for the energy $E_{\rm cm} = 9.21$ MeV.
}
\label{fdpot-phase}
\end{figure}

The difference could be understood from the change of the shapes of the 
frozen density potential energies with respect to the relative distance  
$R$ when the gauge angle $\chi$ is varied. 
In Fig. \ref{fdpot-phase}, we plot the frozen density potential energy curves 
near the point $R =$ 9 fm with the gauge angles $\chi =$ 0, 45, 90, 135 degrees.
We see that the top of the energy curve shifts to the direction of the 
small value of the $R$ when the gauge angle is varied as $\chi =$ 0, 45, 90 degrees.
This is consistent with the shift of the point of the $R$ 
at which the relative momentum $P_{z}$ takes the minimum absolute value 
in the case of the energy $E_{\rm cm}$ = 11.41 MeV.
Together with the shift of the tops of the frozen density potential energy 
curves, they are shifted upward when the gauge angle is 
varied as $\chi =$ 0, 45, and 90 degrees.   
Then, the section point of the potential energy curve with the energy 
$E_{\rm cm} =$ 9.21 MeV shifts toward the large value of the $R$.
This is consistent with the shift of the turning points of the trajectories 
in Fig. \ref{rz-pz-4.8-2} with the angles $\chi =$ 0, 45, and 90 degrees.

To understand this phenomena, we can look at the total energy 
in the frozen calculation that is directly related to the nucleus-nucleus potential 
in Fig. \ref{fdpot-phase}. 
Because the normal density is not affected by the relative phase, 
only the pairing part of the total energy changes 
with the choice of the relative gauge angle,  
\begin{eqnarray}
	E_{\rm pair} = \frac14 \sum_{\alpha \beta \gamma \delta} 
      \overline{v}_{\alpha \beta \gamma \delta} 
                   \kappa^*_{\alpha \beta} \kappa_{ \gamma \delta } .
\end{eqnarray}  
The $\kappa$ matrix can be decomposed into a part coming from the left nuclei 
and a part coming from the right. 
Because the phase of the $\kappa$ coming from the right part is shifted 
by a phase $2 \chi$, the total $\kappa$ matrix is expected to be minimum 
for the angle $\chi=90$ degrees. 
This phase dependence of the potential energy 
illustrates that the nucleus-nucleus interaction energy for $\chi=90$ degrees 
is smaller than that for $\chi=0$. 
Here, the nucleus-nucleus interaction energy is the difference between 
the Coulomb energy and the frozen density potential energy at each point of the 
relative distance $R$.

\begin{figure}[htb]
\begin{center}
\resizebox{0.4\textwidth}{!}{%
  \includegraphics{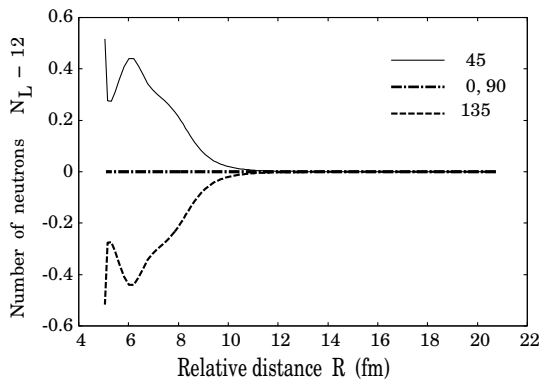}
}
\end{center}
\caption{Variation of the number $N_{\rm L}$ of the neutrons 
in the left region with respect to the 
relative distance $R$ with the energy $E_{\rm cm} = 11.41$ MeV. 
The thin solid (broken) curve is for the case with the gauge angle $\chi =$ 45 (135) 
degrees, respectively.
The chain curve is for the cases with $\chi =$ 0 and 90 degrees. 
The two curves of the cases with $\chi =$ 0 and 90 degrees overlap each other.
}
\label{transfer-phase-hi}
\end{figure}

\begin{figure}[htb]
\begin{center}
\resizebox{0.4\textwidth}{!}{%
  \includegraphics{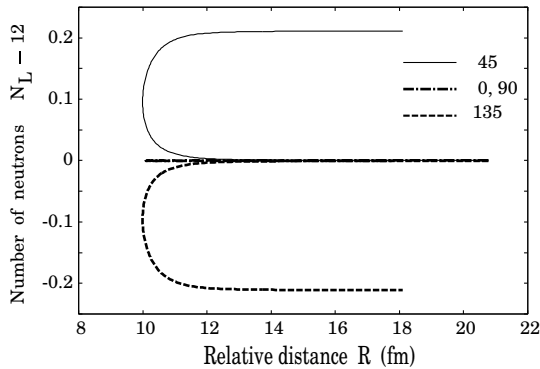}
}
\end{center}
\caption{The same as in Fig. \ref{transfer-phase-hi}, 
but for the energy $E_{\rm cm} =$ 9.21 MeV. 
}
\label{transfer-phase-lo}
\end{figure}

Let us pay attention to the behaviors of the variations of the 
number of the neutrons in the left region. 
The left and right regions are separated by 
a section plane located at $z = z_s$ that is introduced in Section III. 

In Figs. \ref{transfer-phase-hi} and \ref{transfer-phase-lo}, 
we show the variations of the numbers of the neutrons
with respect to the relative distance $R$ when the gauge 
angle is varied as $\chi =$ 0, 45, 90, and 135 degrees.
Each of the curves in the figures stands for the difference of the number 
$N_{\rm L}$ of the neutrons in the left region 
from the initial value 12.
The $N_{\rm L}$ is calculated just like Eq. (\ref{eq:m_L}) 
with $\rho({\bf r})$ replaced by the neutron density $\rho^{(n)}({\bf r})$, 
\begin{eqnarray}
 N_{\rm L} =
     \int d^3 x \,\, \rho^{(n)}({\bf r}) \theta(z_{\rm s} - z)  \, .  \label{eq:N_L} 
\end{eqnarray}  

In both of the figures, the curves for the gauge angles $\chi =$ 0 
and 90 degrees are flat, denoting that the numbers 
of the neutrons in the left region
are kept to be 12 all along the trajectories.
In the cases of the trajectories with the gauge angles $\chi =$ 45 and 
135 degrees, on the other hand, the number of the neutrons 
increase (decrease) for the gauge angle $\chi =$ 45 (90) degrees 
in the left region, respectively.

\begin{figure}[htb]
\begin{center}
\resizebox{0.4\textwidth}{!}{%
  \includegraphics{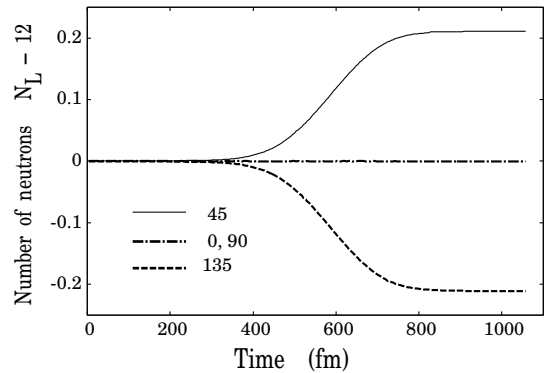}
}
\end{center}
\caption{The same as in Fig. \ref{transfer-phase-lo},  
but the variations of the number $N_{\rm L}$ of the neutrons 
in the left region are plotted with respect to the elapsed time.
}
\label{transfer-time-phase}
\end{figure}

In Fig. \ref{transfer-phase-lo}, we see that the change of the number 
$N_{\rm L}$ of the neutrons is realized near the turning point 
of the trajectory of the bouncing nuclei. 
When the number $N_{\rm L}$ of the neutrons is plotted 
with respect to the elapsed time in Fig. \ref{transfer-time-phase}, 
the smooth variation of the number $N_{\rm L}$ is clearly illustrated. 
Thus, the figures \ref{transfer-phase-hi} to \ref{transfer-time-phase} 
suggest us the dependence of the number $N_{\rm L}$ on the 
relative gauge angle $\Phi = 2 \chi$ with the periodicity 2$\pi$.

Taking account of the one-dimensional situation of the present calculations 
of the head-on collisions, and assuming the 2$\pi$ periodicity of the 
number $N_{\rm L}$ of the neutrons in the left region with respect to the 
relative gauge angle $\Phi$, we guess that the flux $J_{\rm s}$ 
of the neutrons across the section plane at $z = z_{\rm s}$ 
is proportional to the sine of the angle $\Phi$, 
\begin{eqnarray}
  J_{\rm s} \propto \sin \left( \Phi \right).   \label{JsPhi-2}
\end{eqnarray}  

The relation in (\ref{JsPhi-2}) reminds us the Josephson current 
of the electron pair through the Josephson junction 
with a relative phase $\Phi$ of the two superconducting objects 
separated by a thin insulator~\cite{Joseph-1,Joseph-2}.
The detailed studies are needed to understand the microscopic background 
of the flows of the nucleons under the condition of the superfluidity 
illustrated in the figures \ref{transfer-phase-hi} to \ref{transfer-time-phase}.

\section{Summary and concluding remarks}
In this article, we have reported the first results of the application 
of the method of solving the TDHFB equations with the Gogny force 
to the head-on collision processes 
of the superfluid nuclei ${}^{20}$O + ${}^{20}$O.
The method of solving the TDHFB equations are realized by using the two-dimensional 
harmonic oscillator eigenfunctions and one-dimensional Lagrange mesh~\cite{YH-PRC88}.
A candidate of the way of setting up the initial conditions 
of the collision processes was proposed. 

The advantage of the present numerical method is the natural cut-off 
for quasi-particle energy obtained by using 
a Gaussian type finite range interaction. 
Furthermore, the choice of the present hybrid basis allows to include 
all the quasi-particle states, then the unitarity relations 
\eqref{unitarity-1} to \eqref{unitarity-4} are respected during the evolution. 
These two points are expected to contribute to the stable numerical integrations 
of the TDHFB equations.

Setting the energies of the colliding nuclei around the energy of the top 
of the frozen density potential energy, we have displayed the trajectories, 
variations of the pairing energies with respect to the relative distance, 
and the numbers of the transferred nucleons.

We  studied the effects of the initial relative phase 
on the properties of the colliding superfluid nuclei.  
In particular, the dependence of the trajectories, pairing energies, 
and transferred number of the neutrons on the relative phase 
was visually illustrated in the figures.
We showed that the static and dynamical nucleus-nucleus potential
depends on the relative gauge angle between two superfluid nuclei. 
For the  reaction ${}^{20}$O + ${}^{20}$O  the difference 
concerning the potential energy is about 0.4 MeV for the height 
of the barrier and 0.2 fm for the position of the barrier.

Some remarks concerning the present calculations are as follows:\\
1. The set up the initial conditions of the TDHFB equations 
of the colliding superfluid nuclei was of a {\it sharp cut-off} type, 
by assuming that the correlations of the two nuclei at the initial time 
could be neglected when they were set apart with large distance. 
At present, the effects of the sharp cut-off initial conditions  
have not been studied. 
It would be interesting to compare the results 
of the present calculations with those obtained by using other types 
of initial conditions. 

2. 
In all of the numerical calculations in this article, 
we have used a fixed set of the parameters of the basis functions 
(number of the grid points $N_{\rm grid}$, 
grid size $\Delta z$, maximum number of the two-dimensional 
harmonic oscillator shell $N_{\rm shell}$, and harminic oscillator frequency 
$\hbar \omega$, and so on). 
The dependence of the numerical results on the parameters of the basis functions 
will be studied before the heavier nuclei are treated in the future calculations. 

3. In calculating the numbers of the transferred nucleons, we followed 
Washiyama's method proposed in the TDHF framework. 
It would be necessary to study the present results 
from the quantum mechanical viewpoint by making use of 
the number projection method~\cite{Scamps1}.

4. The present contribution shows that there is a dependence of the observables 
with respect to the initial gauge angle. 
Nevertheless, the two fragments should initially respect the gauge angle symmetry 
and preserve the initial good number of particles in each fragments. 
It would be interesting for future applications to develop a self-consistent 
time-dependent mean-field theory that describes 
the evolution of a quasi-particle state projected on the good number of particles.

The transfer mechanism of the nucleons in the collision processes 
of the superfluid nuclei will be studied in a future analysis 
by combining the present method 
of solving the Gogny-TDHFB equations with the quantum mechanical method of 
the number projection~\cite{Scamps1}.

\begin{acknowledgments}
The authors thank the members of the DFT meeting for the exciting discussions.
One of the authors (Y.~H.) thanks Professor T.~Nakatsukasa 
for discussions and comments. 
G.~S. acknowledges the Japan Society for the Promotion of Science 
for the JSPS postdoctoral fellowship for foreign researchers. 
This work was supported by Grant-in-Aid for JSPS Fellows No. 14F04769. 
This research work is partly supported by results 
of HPCI Systems Research Projects (Project ID hp140010 and hp150081).
Part of the numerical calculations was carried out on SR16000 at YITP 
in Kyoto University. 
Using COMA at the CCS in University of Tsukuba, the numerical calculations 
were partly performed.  

\end{acknowledgments}


\end{document}